\documentclass{appolb}
\usepackage{graphicx}
\usepackage{epsfig}
\usepackage{bm}
\usepackage{amssymb}
\usepackage{mathrsfs}
\usepackage{amsmath}
\usepackage{dsfont}
\usepackage{subfigure}

\begin{document}
\title{Worm Algorithm for Abelian Gauge-Higgs Models
\thanks{Excited QCD - Sarajevo 3-9 February, 2013.  Presented by Y.Delgado}}
\author{Ydalia Delgado, Alexander Schmidt
\address{\small Institut f\"ur Physik, Karl-Franzens Universit\"at, Graz, Austria}}
\maketitle

\begin{abstract}
We present the surface worm algorithm (SWA) which is a generalization of the Prokof'ev Svistunov 
worm algorithm to perform the simulation of the dual representation (surfaces and loops) of 
Abelian gauge-Higgs models on a lattice. We compare the SWA to a local Metropolis update in the dual 
representation and show that the SWA outperforms the local update for a wide range of parameters. 
\end{abstract}

\section{Introduction}
\vspace{-1mm}
\noindent 
The complex fermion determinant at finite chemical potential
has slowed down the progress in the exploration of the QCD 
phase diagram using Lattice QCD.
Among the different techniques to deal with the sign problem 
(see e.g. \cite{reviews}), 
the dual representation is a powerful method which can help us to solve the sign problem 
without making any approximation of the partition sum 
as in other methods.
Before approaching the ultimate goal of finding a dual representation for non-Abelian gauge theories
with only positive probability weights, which is a rather involved task and 
has not been achieved yet,
it is advisable to explore and understand the method in
simpler models, such as
Abelian theories coupled to scalar fields \cite{endres,dualz3_ref} that we study here.

Once the partition sum with real and positive probability weights is found, 
the next step is to choose the most efficient algorithm to save computer time.
In the case of only matter fields or spins, 
the worm algorithm \cite{worm} constitutes one of the most suitable and efficient 
methods \cite{worm_critical_behavior} to deal with the constrained degrees of freedom of the system, i.e. with loops.
In this article we present an extension of the worm algorithm (SWA) \cite{swa} to perform 
the simulation of the U(1) gauge-Higgs model where loops and surfaces are the dual variables.
We assess the performance of the SWA in comparison to a local Metropolis update (LMA).
The analysis of the physics of this model will be presented elsewhere \cite{u1}.

\section{Dual representation of an Abelian gauge-Higgs model}
\vspace{-1mm}
\noindent
The action of the U(1) gauge-Higgs model on the lattice is given by
\begin{eqnarray}
 S_G\ & \!=\! & -\frac{\beta}{2} \sum_x \sum_{\nu < \rho} 
\left[ U_{x,\nu\rho}\ +\ U^*_{x,\nu\rho} \right] \nonumber \\
 S_H\ & \!=\! & \sum_{x}\left[\kappa |\phi_x|^2 + \lambda|\phi_x|^4 \right] 
       - \sum_{x,\nu} \left[
     \phi^*_x \, U_{x,\nu}   \phi_{x+\hat{\nu}}
\ +\ \phi_x^* \, U_{x-\hat{\nu},\nu}^* \phi_{x+\hat{\nu}} \right]
\end{eqnarray}
where $U_{x,\nu} =  e^{iA_\nu}\ \in\ U(1),\ A_\nu \in [-\pi,\pi]$ are the link variables, 
$S_G$ is the usual plaquette action, with 
$U_{x,\nu\rho}  =  U_{x,\nu}U_{x+\hat{\nu},\rho}U_{x+\hat{\rho},\nu}^*U_{x,\rho}^*$
the plaquette variable.
The Higgs fields $\phi_x\  \in\  \mathds{C}$ in the Higgs action $S_H$ live on the sites of the lattice.
$\kappa$ is a mass parameter and  $\lambda$ is the quartic coupling.  

\vspace*{3mm}
\noindent 
Here we outline the general strategy for the derivation of the dual representation (for 
the details see the appendix in \cite{swa}). The general steps are:
Write the Boltzmann weight in a factorized form and expand 
the exponentials for individual plaquettes and links.
A single nearest neighbor term turns into 
$$e^{\phi_x^* U_{x,\nu} \phi_{x+\hat{\nu}} } = 
\sum_{l_{x,\nu}} \frac{1}{l_{x,\nu}!} 
(U_{x,\nu})^{l_{x,\nu}} (\phi^*_x)^{l_{x,\nu}} (\phi_{x+\hat{\nu}})^{l_{x,\nu}}\; .$$
A single plaquette term leads to
$$e^{\beta U_{x,\nu} U_{x+\hat{\nu},\rho} U^*_{x+\hat{\rho},\nu} U^*_{x,\rho} } = 
\sum_{p_{x,\nu\rho}} \frac{\beta^{p_{x,\nu\rho}}}
{p_{x,\nu\rho}!} 
\left[U_{x,\nu} U_{x+\hat{\nu},\rho} U^*_{x+\hat{\rho},\nu} U^*_{x,\rho}\right]^{p_{x,\nu\rho}}\;.$$

\noindent 
After integrating out the U(1) variables, 
the new form of the partition sum depends only on the dual variables:
The constrained link occupation number  $l_{x,\nu} \in (-\infty,+\infty)$,
the unconstrained link occupation number $\overline{l}_{x,\nu} \in [0,+\infty)$ and
the constrained plaquette occupation number $p_{x,\nu\rho} \in (-\infty,+\infty)$.
The new form of the partition sum is
\begin{equation}
Z \; \propto \; \sum_{\{p,l,,\overline{l}\}} \; {\cal W} [p,l,\overline{l}] \; {\cal C}_S[l] \; {\cal C}_L[p,l] \;,
\end{equation}
where the new degrees of freedom are the dual variables $l$, $\overline{l}$ and $p$
and $\sum_{\{p,l,,\overline{l}\}}$ denotes the sum over all their configurations.
${\cal W} [p,l,\overline{l}]$ is a positive weight factor. Furthermore, constraints appear
that force the total sum of the occupation numbers to vanish at every site and link:
${\cal C}_S[l]$ is the site constraint which forces the total matter flux to vanish at every site 
and gives rise to loops,
\vspace*{-1mm}
$$
\forall x :\; \sum_{\nu=1}^4 [l_{x,\nu} - l_{x-\hat{\nu},\nu}] = 0 \;.
$$
The link constraint ${\cal C}_L[p,l]$ gives rise to gauge surfaces,
\vspace*{-1mm}
$$
\forall x,\nu:\; \left( \sum_{\rho:\nu<\rho}[p_{x,\nu\rho}
- p_{x-\hat{\rho},\nu\rho}] - \sum_{\rho:\nu>\rho}[ p_{x,\rho\nu}
- p_{x-\hat{\rho},\rho\nu} ] + l_{x,\nu}\right) = 0 \; .
$$

\section{Monte Carlo simulation}
\begin{figure}[h]
\begin{center}
\includegraphics[width=0.7\textwidth,clip]{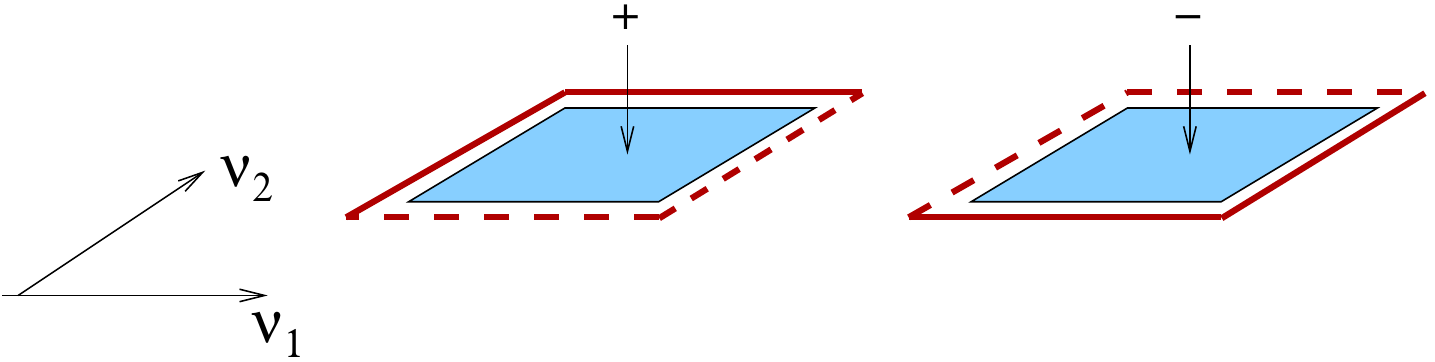}
\end{center}
\vspace{-4mm}
\caption{Plaquette update: A plaquette occupation number is changed by $+1$ (lhs.\ plot) or
$-1$ (rhs.) and the links of the plaquette are changed simultaneously. The
full line indicates an increase by +1 and a dashed line a decrease by $-1$. 
The directions $1 \le \nu_1 < \nu_2 \le 4$
indicate the plane of the plaquette.} \label{plaquette}
\vspace{-2mm}
\end{figure}

\begin{figure}[h]
\begin{center}
\includegraphics[width=0.7\textwidth,clip]{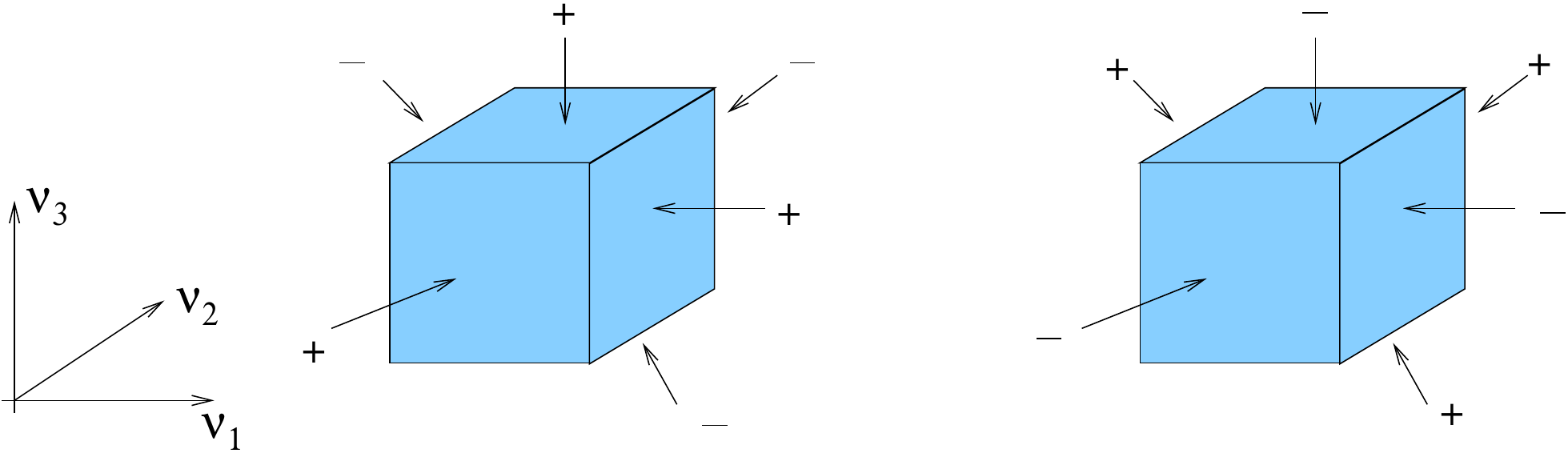}
\end{center}
\vspace{-4mm}
\caption{Cube update: Here we show the changes in the plaquette occupation numbers. 
The edges of the 3-cube are parallel to 
the directions $1 \leq \nu_1 < \nu_2 < \nu_3 \leq 4$.} \label{cube}
\vspace*{-4mm}
\end{figure}

\vspace*{-1mm}
\noindent
To perform the Monte Carlo simulation of the system we developed the SWA
and we compared its performance against a local update (LMA) \cite{dualz3_ref,swa}.
The LMA consists of:
\begin{itemize}
\vspace*{-1mm}
\item A sweep of the unconstrained variables $\overline{l}$ rising or lowering their occupation number by one unit.
\vspace*{-1mm}
\item ``Plaquette update'': 
It consists of increasing or decreasing a plaquette occupation number
$p_{x,\nu\rho}$ and
the link fluxes $l_{x,\sigma}$ at the edges of $p_{x,\nu\rho}$ by $\pm 1$ as 
illustrated in Fig.~\ref{plaquette}. The change of $p_{x, \nu \rho}$ 
by $\pm 1$ is indicated by the signs $+$ or $-$, while for the flux variables we use a dashed line to indicate a decrease by $-1$ and a full line for an increase by $+1$.
\vspace*{-1mm}
\item ``Cube update'':  The plaquettes of 3-cubes
of our 4d lattice are changed according to one of the two patterns illustrated in 
Fig.~\ref{cube}. 
Although the plaquette update is enough to satisfy ergodicity, 
the cube update helps for decorrelation in the region of 
parameters where the link $l$ acceptance rate is low
and the system is dominated by closed surfaces.
\end{itemize}
\vspace*{-1mm}
A full sweep consists of visiting the $4V_4$ links, $6V_4$ plaquettes and $4V_4$ 3-cubes,
offering one of the changes mentioned above and accepting them with the Metropolis 
probability computed from the local weight factors.

\begin{figure}[h]
\begin{center}
\includegraphics[width=\textwidth,clip]{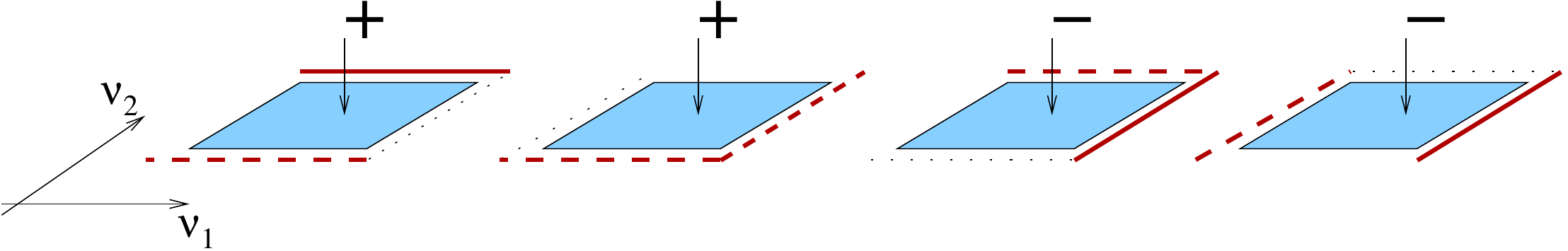}
\end{center}
\vspace{-4mm}
\caption{Examples of positive (lhs.) and negative segments (rhs.) 
in the $\nu_1$-$\nu_2$-plane ($\nu_1 < \nu_2$).
The plaquette occupation numbers are changed as indicated by the signs. 
The full (dashed) links are changed by $+1$ ($-1$). The empty link shows
where the segment is attached to the worm and the dotted link is the new position of the link
$L_V$ where the constraints are violated.} \label{segments}
\vspace{-4mm}
\end{figure}

\begin{figure}[h]
\begin{center}
\includegraphics[width=\textwidth,clip]{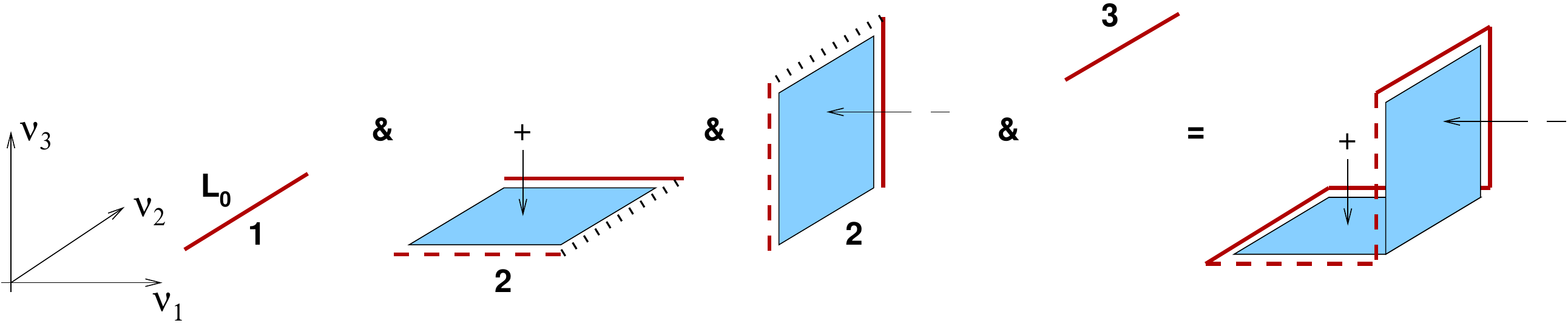}
\end{center}
\vspace{-4mm}
\caption{Illustration of the worm algorithm.  See text for an explanation.} \label{worm}
\vspace{-4mm}
\end{figure}

\noindent
Instead of the plaquette and cube updates we can use the worm algorithm.
The SWA (see \cite{swa} for a detailed description) 
is constructed by breaking up the plaquette update 
into smaller building blocks called ``segments'' 
(examples are shown in Fig.~\ref{segments}) used to grow surfaces  
on which the flux and plaquette variables are changed.
In the SWA the constraints are temporarily violated at a link
$L_V$, the head of the worm, and the two sites at its endpoints.
The admissible configurations are produced using 3 steps: 
The worms starts by changing the flux by $\pm 1$ at a randomly chosen link $L_0$  
(step 1 in Fig.~\ref{worm}). $L_0$ becomes the head of the worm $L_V$.
The defect at $L_V$ is then propagated through the lattice by 
attaching segments, which are chosen in such a way that the constraints are always 
obeyed (2 in Fig.~\ref{worm}). 
The defect is propagated through the lattice until the worm decides to
end with the insertion of another unit of link flux at $L_V$ (3 in Fig.~\ref{worm}).
A full sweep with the SWA consists of $V_4$ worms.

\section{Numerical analysis}
\begin{figure}[t]
\begin{center}
\includegraphics[width=0.9\textwidth,clip]{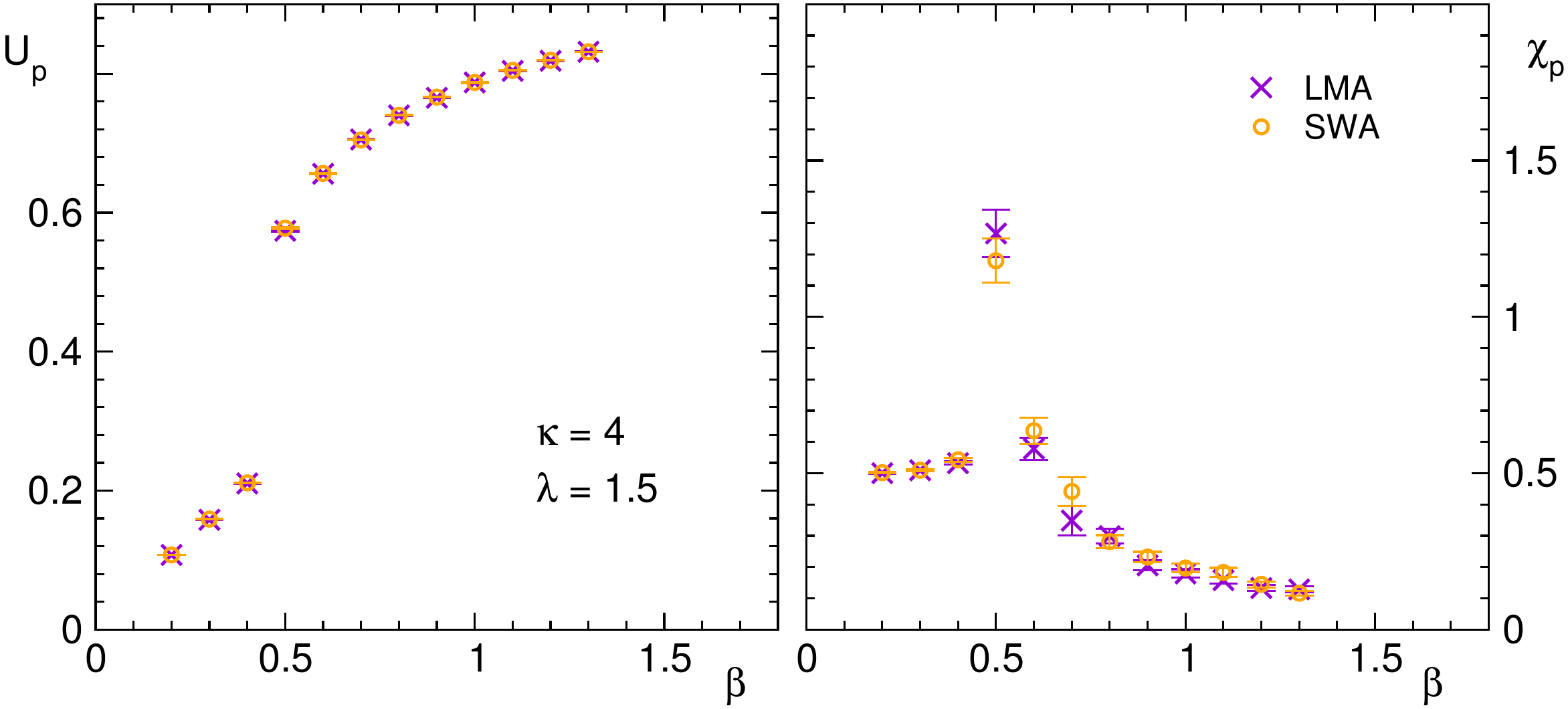}
\includegraphics[width=0.95\textwidth,clip]{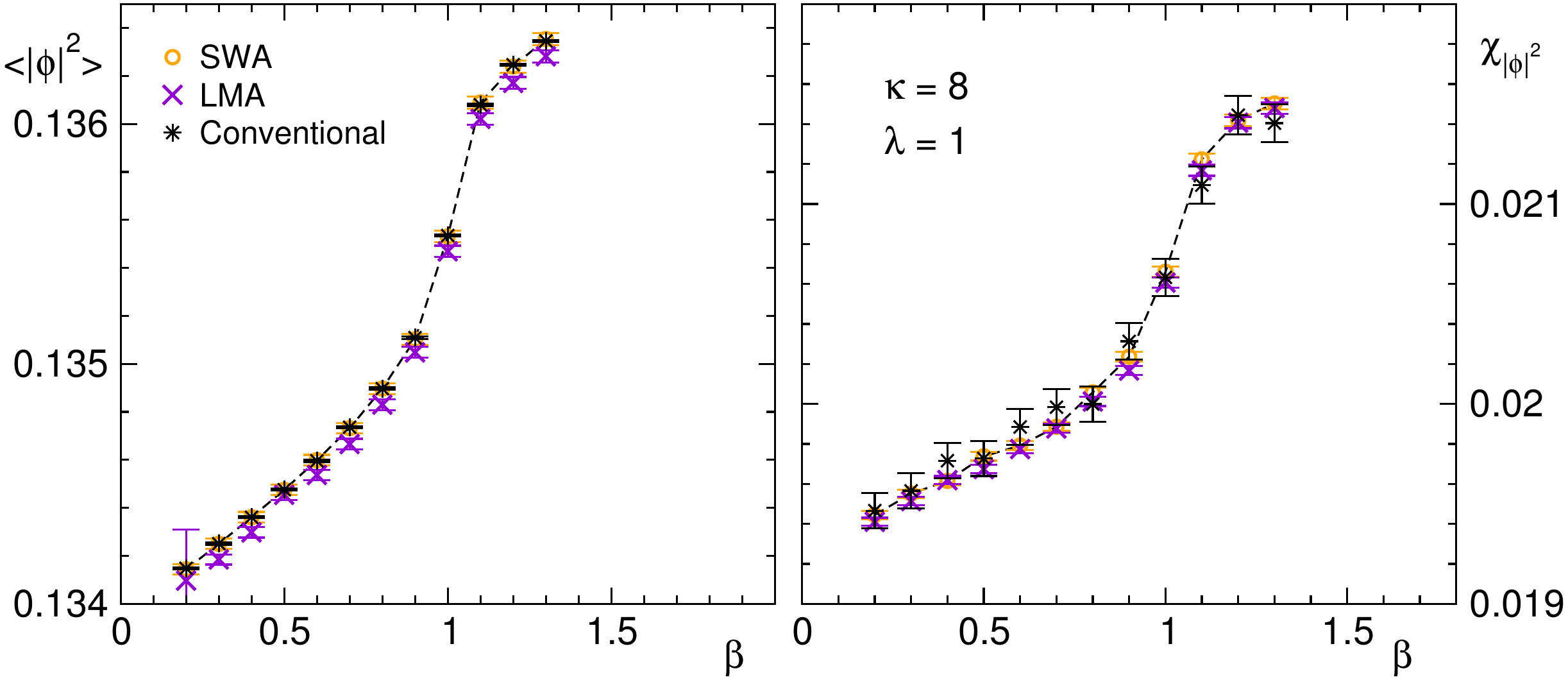}
\end{center}
\vspace{-4mm}
\caption{Observables as a function of $\beta$ for different parameters and volumes.
We compare results from three algorithms: The conventional approach (asterisks), 
the SWA (circles) and the LMA (crosses).} \label{obs}
\vspace*{-4mm}
\end{figure}

\vspace{-1mm}
\noindent
For the comparison of both algorithms we analyzed the bulk observables (and their 
fluctuations): $U_P$ which is the derivative wrt. $\beta$ and $|\phi|^2$ (derivative wrt. 
$\kappa$).  First we checked the correctness of the SWA comparing the results for different 
lattices sizes and parameters.  For example, the upper plot of Fig.~\ref{obs} 
shows $U_P$ as a function of $\beta$ for $\kappa = 4$, $\lambda = 1.5$ on a lattice of size $4^4$.  
The lower plot shows $\langle|\phi|^2\rangle$ for $\kappa = 8$ and $\lambda = 1$ on a $10^4$ lattice.  
In both cases we used $10^6$ equilibration sweeps, $10^6$ measurements and $10$ sweeps 
for decorrelation between measurements.  We observe very good agreement among
the different algorithms.

\begin{figure}[t]
\begin{center}
\includegraphics[width=\textwidth,clip]{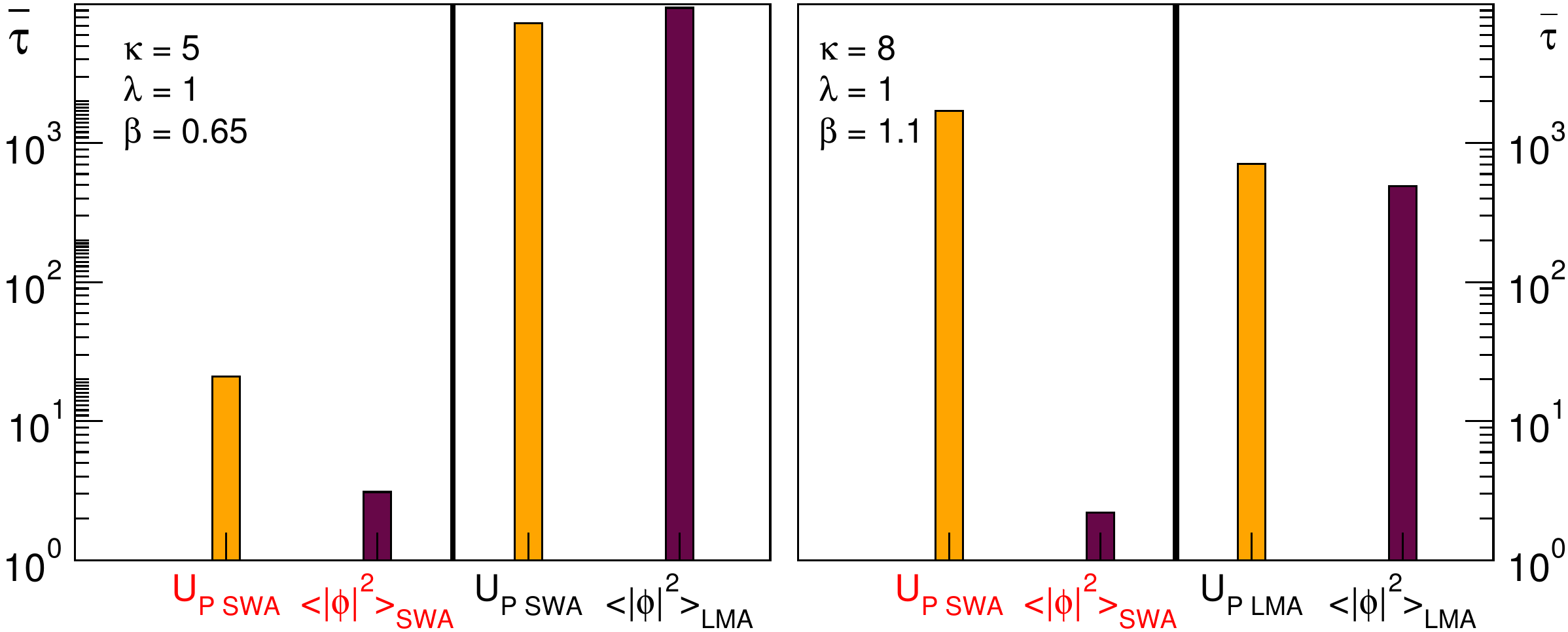}
\end{center}
\vspace{-4mm}
\caption{Normalized autocorrelation times $\overline{\tau}$ for 2 different set
of parameters.  Left: parameters close to a first order phase transition. 
Right: low acceptance rate of the variable $l$.  Both simulations correspond
to a $16^4$ lattice.  Data taken from \cite{swa}.} \label{auto}
\vspace*{-4mm}
\end{figure}

\noindent
In order to obtain a measure of computational effort, we compared the normalized 
autocorrelation time $\overline{\tau}$ as defined in \cite{swa} of the SWA and LMA for 
different volumes and parameters.  We concluded that,
the SWA outperforms the local update near a phase transition and if
the acceptance rate of the link variable $l$ is not very low (eg. lhs. of Fig.~\ref{auto}).  On the other hand, when the links
become expensive the worm algorithm has difficulties to efficiently sample the 
system (as can be observed on the rhs. of Fig.~\ref{auto}, $\overline{\tau}$ for
$U_P$ is larger for the SWA than for the LMA).  But this can be overcome by offering
a sweep of cube updates or a worm made of only plaquettes as described in \cite{endres}.

\section*{Acknowledgments} 
\vspace{-1mm}
\noindent
We thank Hans Gerd Evertz and Christof Gattringer for fruitful discussions at various stages of this work. 
This work was supported 
by the Austrian Science Fund, FWF, DK {\it Hadrons in Vacuum, Nuclei, and Stars} 
(FWF DK W1203-N16)
and by the Research Executive Agency (REA) of the European Union 
under Grant Agreement number PITN-GA-2009-238353 (ITN STRONGnet).

\end{document}